\documentclass[twocolumn,showpacs,preprintnumbers,amsmath,amssymb]{revtex4}

\usepackage{graphicx}
\usepackage{dcolumn}
\usepackage{bm}

\begin{document}


\title{Photoluminescence investigations of 2D hole Landau levels\\
 in p-type single Al$_{x}$Ga$_{1-x}$As/GaAs heterostructures}

\author{M. Kubisa}
\email{kubisa@if.pwr.wroc.pl}
\author{L. Bryja, K. Ryczko, J. Misiewicz}
\affiliation{Institute of Physics, Wroc{\l}aw University of Technology,\\
Wybrze{\.z}e Wyspia{\'n}skiego 27, 50-370 Wroc{\l}aw, Poland}%

\author{C. Bardot, M. Potemski}
\affiliation{ Grenoble High Magnetic Field Laboratory MPI-FKF/CNRS, 25 Avenue des Martyrs, F-38042 Grenoble Cedex 9, France}%

\author{G. Ortner, M. Bayer, A. Forchel}
\affiliation{Technische Physik, Universit{\"a}t W{\"u}rzburg, Am Hubland, 97074 W{\"u}rzburg, Germany}%

\author{C. B. S{\o}rensen}
\affiliation{ The Niels Bohr Institute, University of Copenhagen, Universitetsparken 5, DK-2100 Copenhagen, Denmark}%

\date{\today}

\begin{abstract}
We study the energy structure of two-dimensional holes in \textit{p}-type single 
Al$_{1-x}$Ga$_{x}$As/GaAs heterojunctions under a perpendicular magnetic field. Photoluminescence measurments with low densities of excitation power reveal rich spectra containing both free and bound-carrier transitions. The experimental results are compared with energies of valence-subband Landau levels calculated using a new numerical procedure and a good agreement is achieved. Additional lines observed in the energy range of free-carrier recombinations are attributed to excitonic transitions. We also consider the role of many-body effects in photoluminescence spectra.
\end{abstract}

\maketitle




\section{INTRODUCTION}
Valence band subbands in doped quantum wells and heterojunctions have complicated structure. Due to strong electric fields resulting from both band-edge discontinuities and irregular charge distribution the heavy and light-hole states are mixed. This results in a strong nonparabolicity and anisotropy of energy levels and their nonlinear behavior under an external magnetic field. Hole subbands in doped heterostructures were studied experimentally, however much less intensively than conduction band states. In magnetotransport measurements \cite{Stor1,Eisen2,Hans3,Kemer4,Kolok5} effective masses of holes were determined from Shubnikov-De Haas oscillations. It was shown \cite{Stor1,Eisen2} that in single heterojunctions the spin degeneracy of valence band subbands is lifted as a result of the asymmetry of confining potential and the large spin-orbit coupling. The complex arrangement of hole Landau levels was investigated in cyclotron resonance experiments \cite{Stor1,Schles6,Erhar7,Iwasa8,Schles9}. However the photoluminescence (PL) spectroscopy in a magnetic field normal to the layers, very effective in providing information on the conduction band electrons, was relatively seldom used for \textit{p}-doped structures \cite{Yuan10,Ossau11,Silov12,Butov13,Grav14,Volk15,Shen16}. Due to higher effective masses both the cyclotron energy and mobility are smaller in the valence band by nearly an order of magnitude. As a result the ratio of Landau level broadening (increasing with diminishing mobility) to their distance becomes substantially larger, reducing the resolution of measurements. Usually the studied luminescence resulted from band-to-band transitions. Volkov et al. \cite{Volk15} investigated radiative recombination of two-dimensional (2D) holes with donor-bound electrons in specially designed heterojunctions where a monolayer of donors was introduced at some distance from the interface. In recent publications \cite{Finke17,Ponom18,Glas19} observations of subtle many-body effects like positively charged excitons or shake-up processes were reported in \textit{p}-type quantum wells. To reduce the broadening of PL features in these experiments free holes were created not by means of a modulation doping but using optical excitation. \\
\hspace*{0.2cm} Calculations of hole Landau levels in \textit{p}-doped heterostructures are quite complex. Similarly as in the case of conduction-band subbands one has to solve self-consistently Schr{\"o}dinger and Poisson's equations. But valence-band states are described by the 4$\times$4 Luttinger Hamiltonian \cite{Lutt20} which considerably complicates the computations. In early attempts \cite{Ando21,Broi22,Eken23,Yang24,Bang25} the hole eigenstates were expanded in a limited set of basis functions and calculated by matrix diagonalization. The precision of such methods strongly depends on the choice of the basic set and is known to fail for larger magnetic fields unless the number of basis functions is increased \cite{Ando21}. Bangert and Landwehr \cite{Bang25} examined the importance of cubic anisotropy terms commonly ignored in calculations. Theory of exchange-correlation effects in the valence band was developed by Bobbert et al. \cite{Bobb26} who proposed a simple method of including many-body interactions in subband calculations. In Refs \cite{Volk15} and \cite{Kemer27, Kemer27A} improved numerical algorithms for solving the Luttinger Hamiltonian were elaborated. The structure of hole Landau levels calculated by different authors was compared with experimental results, however the agreement was rather poor. \\
\hspace*{0.2cm} In this paper we investigate low-temperature photoluminescence from \textit{p}-type single Al$_{1-x}$Ga$_{x}$As/GaAs heterostructures in a magnetic field perpendicular to the layers. As it was first demonstrated by Yuan et al. \cite{Yuan28,Yuan28A} the PL spectrum of a modulation-doped heterostructure is dominated by a broad line, named H-band, located in the energy range of bulk excitons and donor-acceptor pairs recombination. The characteristic feature of this line, observed both in donor and acceptor-doped structures, is a strong blue shift with increasing excitation power. A tilted magnetic field experiment \cite{Ossau11} proved that 2D carriers are involved in the H-band emission. Time resolved PL measurements \cite{Berg29} displayed extremely long life times which strongly depended on the line position. This was explained as a result of spatial separation (by the interface electric field) of carriers participating in the recombination process. Along with these observations the H-band emission was interpreted as a recombination of 2D particles confined near the junction (electrons or holes according to the doping) with photocreated carriers of the opposite charge removed by the electric field far from the interface. In strongly doped heterostructures with high densities of 2D carriers the large interface electric field results in appreciable spatial separation of recombining electrons and holes. As a consequence the H-band emission has a free-carrier character which reveals in linear changes of the line position with a perpendicular magnetic field \cite{Ossau11,Shen16,Kim30,Bryja31}. In less doped structures an excitonic nature of the H-band line was observed as a diamagnetic shift of the transition energy \cite{Shen16,Shen32}. This clearly results from reduced separation of recombining particles which enhances their Coulomb binding. \\
\hspace*{0.2cm} Free-carrier recombination observed in our samples with high densities of 2D holes enabled us to derive the structure of valence-subband Landau levels. In PL measurements we applied very low densities of excitation power ($<0.16$ mW/cm$^{2}$), which increased the resolution and allowed us to discern the fine structure of H-band emission. In addition to free-carrier lines we observed a recombination of donor-bound electrons with 2D holes. The sample parameters as well as details of the experimental methods are presented in Sec. II. To analyze the results we calculated the energies and wave functions of 2D hole Landau levels using a new exact numerical procedure that will be described in Sec. III. We assumed that the non-doped GaAs layer is of \textit{p}-type and the 2D holes form an accumulation layer at the interface. Usually (see for example Ref. \cite{Weim33}) the background doping of pure MBE-grown GaAs is of \textit{p}-type in the low $10^{14}$ cm$^{-3}$ range. This results in weaker electric field near the interface and consequently in somewhat lower 2D hole subband energies as compared with the previous calculations, where the \textit{n}-type residual GaAs doping was assumed. We also evaluated the matrix elements for optical transitions between valence and conduction-band Landau levels. The theoretical results are compared with measured energies of PL transitions in Sec. IV and a very good agreement is established. Besides the free and donor-related transitions new lines are discerned in the spectra at higher magnetic fields, possibly of excitonic nature. We discuss the influence of exchange-correlation effects on hole Landau levels and observed recombination processes. Section V will summarize our conclusions.

\section{EXPERIMENTAL SYSTEMS AND SAMPLE PARAMETERS}
We investigated two single \textit{p}-type GaAs/Ga$_{0.5}$Al$_{0.5}$As heterostructures with hole concentrations p$_{1} = 7.6{\times}10^{11}$ cm$^{-2}$ and p$_{2}$ = $9.8{\times}10^{11}$ cm$^{-2}$. They were grown by molecular beam epitaxy on $(001)$ semi-isolating GaAs substrates in the following sequence: $1$ mm undoped GaAs, $7$ nm undoped Ga$_{0.5}$Al$_{0.5}$As spacer, $50$ nm Ga$_{0.5}$Al$_{0.5}$As layer doped with Be $(5{\times}10^{17}$ cm$^{-3}$ for sample $1$ and $1{\times}10^{18}$ cm$^{-3}$ for sample $2)$ and $5$ nm GaAs cap layer also doped with Be $(1{\times}10^{18}$ cm$^{-3}$ for sample $1$ and $2{\times}10^{18}$ cm$^{-3}$ for sample $2$). Transport properties of both structures in helium temperatures were reported previously \cite{Hans3,Kolok5,Hans34}. \\
\hspace*{0.2cm} Photoluminescence measurements were performed using two set-ups. For preliminary studies we used an optical split-coil cryomagnet providing magnetic fields up to $8$ T. The samples were excited in the Faraday configuration with Ar$^{+}$ laser (wavelength $488$ nm). Luminescence spectra were analyzed in $1$ m single grating monochromator
with Nitrogen-cooled CCD camera used as a detector. The second set-up allowed extending PL measurements up to the magnetic field of $14$ T. An optical fiber system was used with the quarter-wave plate and a linear polarizer placed together with the sample in liquid helium. In order to examine the circular $(\sigma^{+}$ and $\sigma^{-})$ polarization of photoluminescence the direction of magnetic field was changed. The spectra were analyzed in the same way as in the first set-up.

\section{THEORY}
To calculate the energy spectrum of 2D holes we developed a technique that can easily be explained for a particle with parabolic dispersion. Suppose that we want to determine bound-state energies and wave functions of the Hamiltonian
\begin{eqnarray}
H=-\frac{\hbar^{2}}{2m(z)}\Delta+V(z),
\label{eq:1}
\end{eqnarray}
in which both the potential $V$ and effective mass $m$ are position dependent only inside some interval $Z_{1}<z<Z_{2}$. By introducing points $z_{0}=Z_{1}<z_{1}<\ldots<z_{N}=Z_{2}$ we divide this interval into $N$ parts and then approximate the actual functions $V(z)$ and $m(z)$ in each subinterval by appropriate constants (such as $V_{j} = V(z_{j})$ and $m_{j} = m(z_{j})$, respectively, for $z_{j-1}<z<z_{j}$). After that local solutions of the Schr{\"o}dinger equation can easily be found:
\begin{eqnarray*}
 F_{j}(\vec{r})= \exp(i\vec{k}_{\bot}\vec{r}_{\bot})f_{j}(z) {\quad} (z_{j-1}<z<z_{j}),
\end{eqnarray*}
\begin{eqnarray}
f_{j}(z)=A_{j}\exp(ik_{j}z)+B_{j}\exp(-ik_{j}z),
\label{eq:2}
\end{eqnarray}
where
\begin{eqnarray}
k_{j}=\sqrt{\frac{2m_{j}}{\hbar^{2}}(E-V_{j})-k_{\bot}^{2}}
\label{eq:3},
\end{eqnarray}
$E$ and $\vec{k}_{\bot}=(k_{x},k_{y})$ are the particle energy and wave vector,
respectively, and $A_{j}$, $B_{j}$ are some unknown coefficients. Equations ~(\ref{eq:2}) and 
~(\ref{eq:3}) are also relevant in the exterior regions $z<z_{0}$ $(j = 0)$ and $z>z_{N}$ $(j = N+1)$. \\
\hspace*{0.2cm} The matching conditions (i.e. the continuity of both the wave function and its derivative divided by the effective mass) at points of division can be handy written in a matrix form. Introducting column vectors $U_{j}=[A_{j};B_{j}]$ we obtain 
\begin{eqnarray}
B_{j-1}(z_{j})U_{j-1}=B_{j}(z_{j})U_{j},{\quad} j= 1,\ldots,N+1
\label{eq:4},
\end{eqnarray}
where
\begin{eqnarray}              
B_{j}(z)= \left[ \begin{array}{cc}
       \exp(ik_{j}z) & \exp(-ik_{j}z) \\
       \frac{ik_{j}}{m_{j}}\exp(ik_{j}z) & -\frac{ik_{j}}{m_{j}}\exp(-ik_{j}z)  
              \end{array} \right].             
\label{eq:5}
\end{eqnarray}
Equations~(\ref{eq:4}) allow to relate the vectors $U_{0}$ and $U_{N+1}$ in the exterior regions:
\begin{eqnarray}
\left[ \begin{array}{c}
       A_{0} \\
       B_{0} 
       \end{array} \right]
= T \left[ \begin{array}{c}
       A_{N+1} \\
       B_{N+1} 
           \end{array} \right]              
\label{eq:6},
\end{eqnarray}
where the transfer matrix $T$ is given by the product
\begin{eqnarray}
T= \prod_{j=1}^{N+1}B^{-1}_{j-1}(z_{j-1})B_{j}(z_{j}).
\label{eq:7}
\end{eqnarray}
For a bound state the wave vectors $k_{0}$ and $k_{N+1}$ must be imaginary as well as the coefficients $A_{0}$ and $B_{N+1}$ have to vanish. According to equation~(\ref{eq:6}) this implies
\begin{eqnarray}
T_{11}(E)= 0.
\label{eq:8}
\end{eqnarray} 
\hspace{0.2cm} To determine the energies of bound states we calculate numerically transfer matrix~(\ref{eq:6}) for different values of the variable $E$ (and a fixed wave vector $\vec{k}_{\bot}$) searching for zeros of the diagonal element $T_{11}$. After that the wave functions can be evaluated by assuming $A_{N+1}=1$ and calculating the coefficients $A_{j}$, $B_{j}$ in successive subintervals from equations~(\ref{eq:4}). The precision of results depends on the density of division (number $N$ of subintervals) and can easily be controlled. If the potential $V(z)$ is not constant for $z\rightarrow\pm\infty$ a good approximation of accurate bound states can be obtained by choosing the outermost points of division ($z_{0}$ and $z_{N}$) suitably far from the classical turning points. \\
\hspace*{0.2cm} This method of calculations can be also applied to matrix Hamiltonians. Holes in the valence band of III-V compounds are described by the Luttinger matrix \cite{Lutt20}
\begin{eqnarray}
H_{L}=- \left[ \begin{array}{cccc}
       P+Q & L & M & 0 \\
       L^{+} & P-Q & 0 & M \\
       M^{+} & 0 & P-Q & -L \\
       0 & M^{+} & -L^{+} & P+Q \\  
       \end{array} \right]             
\label{eq:9},
\end{eqnarray}
where
\begin{eqnarray*}
P=\frac{\hbar^{2}\gamma_{1}}{2m_{0}}K^{2}, 
Q=\frac{\hbar^{2}\gamma_{2}}{2m_{0}}(K^{2}_{x}+K^{2}_{y}-2K^{2}_{z}),
\end{eqnarray*}
\vspace{-0.9cm} 
\begin{eqnarray*} 
L=-i\sqrt{3}\frac{\hbar^{2}\gamma_{3}}{m_{0}}(K_{x}-K_{y})K_{z},
\end{eqnarray*}
\vspace{-0.9cm}
\begin{eqnarray*}   
M=\sqrt{3}\frac{\hbar^{2}\gamma_{2}}{2m_{0}}(K^{2}_{x}-K^{2}_{y})-i\sqrt{3}\frac{\hbar^{2}\gamma_{3}}{m_{0}}K_{x}K_{y}, 
\end{eqnarray*}
\vspace{-0.9cm}
\begin{eqnarray}  
\vec{K}=-i\vec{\nabla},
\label{eq:10}
\end{eqnarray}
$m_{0}$ is the free electron mass and $\gamma_{1}$, $\gamma_{2}$, $\gamma_{3}$ are Luttinger parameters. The eigenenergies of Hamiltonian~(\ref{eq:9}) are
\begin{eqnarray}
\lefteqn{E_{\pm}\big(\vec{k}\big)= -P\big(\vec{k}\big)\pm}
\nonumber\\
& & {}\sqrt{Q^{2}\big(\vec{k}\big)+L\big(\vec{k}\big)L^{+}\big(\vec{k}\big)
+M\big(\vec{k}\big)M^{+}\big(\vec{k}\big)}{},
\label{eq:11}
\end{eqnarray}
where $E_{+}$ and $E_{-}$ relate to light and heavy holes, respectively. The wave functions can be written as $F(\vec{r})= F(\vec{k})exp(i\vec{k}\vec{r})$, where $F(\vec{k})$ is one of the following vectors \cite{Andr35}:
\begin{eqnarray}
L_{1}(\vec{k})=\left[ \begin{array}{c}
                     R_{1} \\
                     L^{+} \\
                     M^{+} \\
                     0
                      \end{array} \right],
L_{2}(\vec{k})=\left[ \begin{array}{c}
                     0 \\
                     M \\
                     -L \\
                     R_{1}
                      \end{array} \right],\nonumber\\      
H_{1}(\vec{k})=\left[ \begin{array}{c}
                     -L \\
                     R_{2} \\
                     0 \\
                     -M^{+}
                      \end{array} \right],
H_{2}(\vec{k})=\left[ \begin{array}{c}
                     -M \\
                     0 \\
                     R_{2} \\
                     L^{+}
                      \end{array} \right].                                                
\label{eq:12}
\end{eqnarray}
Here $R_{1} = Q-P-E_{+}$, $R_{2} = Q+P+E_{-}$, vectors $L_{1}$, $L_{2}$ correspond to light-hole states and vectors $H_{1}$, $H_{2}$ to heavy-hole ones. \\
\hspace*{0.2cm} In heterostructures the Luttinger parameters have different values in subsequent layers. Thus both components of the Hamiltonian $H_{L}+V(z)$ describing hole subbands vary with the distance \textit{z}. 
To find the subband energies we proceed similarly as for Hamiltonian~(\ref{eq:1}). The local solution of Schr{\"o}dinger equation in the subinterval $z_{j-1} < z < z_{j}$, where values of $V$, $\gamma_{1}$, $\gamma_{2}$ and $\gamma_{3}$ are set fixed, can be written in the form
\begin{eqnarray*}
F_{j}(\vec{r})=\exp(i\vec{k}_{\bot}\vec{r}_{\bot})f_{j}(z),
\end{eqnarray*}  
\begin{widetext}
\begin{eqnarray*}
f_{j}(z) = [C_{1}H_{1}(k_{H})+C_{2}H_{2}(k_{H})]\exp(ik_{H}z)+
[C_{3}L_{1}(k_{L})+C_{4}L_{2}(k_{L})]\exp(ik_{L}z)+
\end{eqnarray*}
\begin{eqnarray}
[C_{5}H_{1}(-k_{H})+C_{6}H_{2}(-k_{H})]\exp(-ik_{H}z)+
[C_{7}L_{1}(-k_{L})+C_{8}L_{2}(-k_{L})]\exp(-ik_{L}z).
\end{eqnarray}
\end{widetext}
Here the complex wave vectors $k_{L}$, $k_{H}$ are roots of equations
\begin{eqnarray}
E_{+}(\vec{k}_{\bot},k_{z})=E-V_{j},{\quad}
E_{-}(\vec{k}_{\bot},k_{z})=E-V_{j},
\label{eq:14}
\end{eqnarray}
respectively. We choose $k_{L}$ and $k_{H}$ such that their imaginary parts are nonnegative. Function (13) contains eight unknown coefficients, which will be represented by the column vector $U_{j} =[C_{1};C_{2};\ldots;C_{8}]$. Vectors $U_{j}$ in adjacent subintervals can be related using the matching conditions \cite{Andr35}:
\begin{eqnarray*}
f_{j-1}(z_{j})=f_{j}(z_{j}),
\end{eqnarray*}
\begin{eqnarray}
D_{j-1}f_{j-1}(z_{j})=D_{j}f_{j}(z_{j}),
\label{eq:15}
\end{eqnarray}
where $D_{j}$ is the matrix operator
\begin{widetext}
\begin{eqnarray}
D=- \left[ \begin{array}{cccc}
 (\gamma_{1}-\gamma_{2})\partial/\partial z & \sqrt{3}\gamma_{3}(k_{x}-ik_{y}) & 0 & 0 \\
 -\sqrt{3}\gamma_{3}(k_{x}+ik_{y}) & (\gamma_{1}+\gamma_{2})\partial/\partial z & 0 & 0 \\
 0 & 0 & (\gamma_{1}+\gamma_{2})\partial/\partial z & -\sqrt{3}\gamma_{3}(k_{x}-ik_{y}) \\
 0 & 0 & \sqrt{3}\gamma_{3}(k_{x}+ik_{y}) & (\gamma_{1}-\gamma_{2})\partial/\partial z \\            \end{array} \right]             
\label{eq:16},
\end{eqnarray}
\end{widetext}
with values of Luttinger parameters appropriate for the subinterval $j$. Once more we obtain equations~(\ref{eq:4}), this time with $8\times8$ matrices $B_{j}$ of the form 
\begin{widetext}
\begin{eqnarray}
B_{j}=\left[ \begin{array}{cccc}
 L_{1}(k_{L})\exp(ik_{L}z) & L_{2}(k_{L})\exp(ik_{L}z) & H_{1}(k_{H})\exp(ik_{H}z) & H_{2}(k_{H})\exp(ik_{H}z) \\
 D_{j}L_{1}(k_{L})\exp(ik_{L}z) & D_{j}L_{2}(k_{L})\exp(ik_{L}z) & D_{j}H_{1}(k_{H})\exp(ik_{H}z) & D_{j}H_{2}(k_{H})\exp(ik_{H}z) 
             \end{array} \right. 
\nonumber\\             
\left.  \begin{array}{cccc}
 L_{1}(-k_{L})\exp(-ik_{L}z) & L_{2}(-k_{L})\exp(-ik_{L}z) & H_{1}(-k_{H})\exp(-ik_{H}z) & H_{2}(-k_{H})\exp(-ik_{H}z) \\
 D_{j}L_{1}(-k_{L})\exp(-ik_{L}z) & D_{j}L_{2}(-k_{L})\exp(-ik_{L}z) & D_{j}H_{1}(-k_{H})\exp(-ik_{H}z) & D_{j}H_{2}(-k_{H})\exp(-ik_{H}z) 
             \end{array} \right]                        
\label{eq:17}.
\end{eqnarray}
\end{widetext}

\begin{figure}
\includegraphics[width=8.6cm]{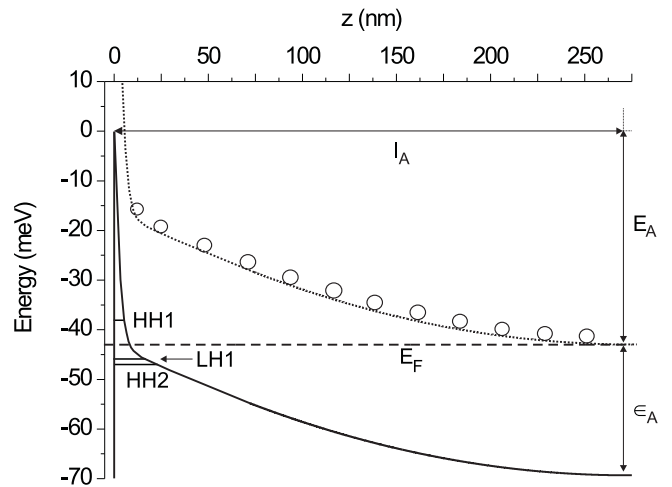} 
\caption{\label{fig:1} Calculated valence-band profile in the well layer of a Al$_{0.5}$Ga$_{0.5}$As/GaAs heterostructure with the 2D-hole density of $7.6{\times}10^{11}$ cm$^{-2}$. HHn and LHn label the edges of heavy and light-hole subbands, respectively. The dotted line represents the acceptor level.}
\end{figure}

To find an equivalent of condition~(\ref{eq:8}) for bound-state energies we examine the structure of vectors $U_{0}$ and $U_{N+1}$ in the exterior regions. They are related by the equation $U_{0} =TU_{N+1}$ with transfer matrix~(\ref{eq:7}) evaluated by means of matrices~(\ref{eq:17}). The complete wave function is evanescent as $z\rightarrow\pm\infty$ only if imaginary parts of both $k_{L}$ and $k_{H}$ are nonzero (i.e. positive according to the definition) for $z<z_{0}$ and $z>z_{N}$. Moreover the components $C_{1},\ldots, C_{4}$ of vector $U_{0}$ must be zero and similarly $C_{5} = C_{6} = C_{7} = C_{8} = 0$ in the case of vector $U_{N+1}$ as can be deduced from expression (13). We thus obtain the equation
\begin{eqnarray}
T_{1} \left[ \begin{array}{c}
            C_{1} \\
            C_{2} \\
            C_{3} \\
            C_{4}
              \end{array} \right]
= \left[ \begin{array}{c}
            0 \\
            0 \\
            0 \\
            0
          \end{array} \right],
\label{eq:18}
\end{eqnarray}
where
\begin{eqnarray}
T_{1}= \left[ \begin{array}{ccc}
       T_{11} & \ldots & T_{14} \\
       \vdots &         & \vdots \\
       T_{41} & \ldots  & T_{44} \\  
       \end{array} \right]                                
\label{eq:19}
\end{eqnarray}
is a $4{\times}4$ submatrix of transfer matrix $T$. Non trivial solutions of equations~(\ref{eq:18}) exist only if
\begin{eqnarray}
\textrm{det}[T_{1}(E)]=0,                       
\label{eq:20}
\end{eqnarray}
which is the requested condition for energies of bound states. To calculate the wave functions we first determine the components $C_{1},\ldots,C_{4}$ of vector $U_{N+1}$ by solving equation~(\ref{eq:18}). After that relations~(\ref{eq:4}) are used to compute vectors $U_{j}$ in succeeding subintervals. Numerical calculations proceed similarly as in the case of scalar Hamiltonian~(\ref{eq:1}) except that higher-order matrices~(\ref{eq:17}) must be evaluated.\\
\hspace*{0.2cm} To account for a magnetic field $\vec{B}$ parallel to $z$-axis operator $\vec{K}$~(\ref{eq:10}) is replaced by  $\vec{K}=-i\vec{\nabla}+e\vec{A}/\hbar c$ and the $\kappa$ terms \cite{Lutt36} are included in Hamiltonian~(\ref{eq:9}). Here $\vec{A}$ is the vector potential, $e$ is the electronic charge and $c$ is the velocity of light. The resulting Hamiltonian can be written in terms of the creation and destruction operators $a^{\pm} = \sqrt{{\hbar}c/2eB}(K_{x}\pm iK_{y})$. We apply the axial approximation \cite{Yang24} by replacing operator $M$~(\ref{eq:10}) by $M=\sqrt{3}\big(\hbar^{2}\bar{\gamma}/2m_{0}\big)(K_{x}-iK_{y})^{2}$, with $\bar{\gamma}=(\gamma_{2}+\gamma_{3})/2$. Then the column vector
\begin{eqnarray}
\Phi_{n}(\dot{r})= \left[ \begin{array}{c}
            F_{1}(z)\phi_{n-1}(\vec{r}_{\bot}) \\
            F_{2}(z)\phi_{n}(\vec{r}_{\bot}) \\
            F_{3}(z)\phi_{n+1}(\vec{r}_{\bot}) \\
            F_{4}(z)\phi_{n+2}(\vec{r}_{\bot})
              \end{array} \right]
\label{eq:21}
\end{eqnarray}
becomes an eigenvector. The harmonic oscillator functions $\phi_{n}$ satisfy $a^{+}\phi_{n}=\sqrt{n+1}\phi_{n+1}$ and $a^{-}\phi_{n}=\sqrt{n}\phi_{n-1}$. In expression~(\ref{eq:21}) the index $n$ runs over the values $n=-2,-1,0,1,\ldots$ and the envelope functions $F_{l}(z)$ are automatically zero for those components which have harmonic oscillator functions $\phi_{n}$ with $n$ negative. By substituting vector~(\ref{eq:21}) to the Schr{\"o}dinger equation $H_{L}F=EF$ we obtain a set of ordinary differential equations for the $z$-dependent components, which can be rewritten as some $4{\times}4$ matrix acting upon $F_{1}(z),\ldots,F_{4}(z)$ functions only. The solutions of this set can easily be obtained with a resemblance to equations~(\ref{eq:11}) and~(\ref{eq:12}). The procedure of transfer matrix evaluation following from this point is analogous to that for the $B=0$ case.\\
\hspace*{0.2cm} In a selectively doped single heterostructure each hole moves in a potential which can be written as a sum
\begin{eqnarray}
V(z)=V_{o}(z)+V_{I}(z)+V_{S}(z).
\label{eq:22}
\end{eqnarray}
Here the component
\begin{eqnarray}
V_{o}(z)= \left\{ \begin{array}{ll}
-V_{o}, & z \leqslant 0, \\
 0,     & z>0,
\end{array} \right.
\label{eq:23}
\end{eqnarray}
describes the valence band offset at the junction ($z=0$). The potential $V_{I}(z)$ created by the immobile charge (ionized impurities) and the potential $V_{S}(z)$ due to mobile 2D holes can be both calculated in the Hartree approximation by solving Poisson's equation
\begin{eqnarray}
\frac{d^{2}V}{dz^{2}}=\frac{4\pi e}{\epsilon}\rho (z),
\label{eq:24}
\end{eqnarray}
with the suitable charge density $\rho(z)$. The dielectric constants $\epsilon$ of GaAs and Ga$_{1-x}$Al$_{x}$As are slightly different, but the resulting image potential was shown \cite{Stern37} to have quite small effect on subband energies. Therefore we neglect it in the calculations. \\
\hspace*{0.2cm} Our structures are doped with acceptors of density $N_{AB}$ in the Ga$_{1-x}$Al$_{x}$As barrier, however a layer of thickness $w$ (the spacer layer) is left undoped to improve  2D carrier mobilities. The holes removed from the acceptor levels form the 2D carrier gas with areal density $N_{S}$ as well as the accumulation layer of width $l_{A}$. We assume that all barrier acceptors are ionized for  $-l_{I} < z < -w$ and similarly all GaAs donors for $0 < z < l_{A}$. Then the resulting immobile charge density is
\begin{eqnarray}
\rho_{I}(z)= \left\{ \begin{array}{ll}
-eN_{AB} & \big(-l_{I} < z < -w\big) \\
+eN_{D}  & \big(0 < z < l_{A}\big)
\end{array} \right.
\label{eq:25},
\end{eqnarray}
where $N_{D}$ is the concentration of residual (minority) donors. Residual acceptors with concentration $N_{A}$ $(>N_{D})$ are neutral in the accumulation layer. The condition of charge conservation can be written as:
\begin{eqnarray}
N_{AB}l_{I}=N_{S}+N_{D}l_{A}.
\label{eq:26}
\end{eqnarray}
Calculating the density of mobile charge we assume that only the ground valence subband is occupied by holes. As we will see this assumption is fulfilled even in our highly doped structures. We also neglect a coupling of the hole motion in the z direction and the motion in the plane of junction, assuming that all 2D holes have the same envelope function $f(z)$. Then the density of mobile charge is
\begin{eqnarray}
\rho_{S}(z)=eN_{S}f^{2}(z).
\label{eq:27}
\end{eqnarray}
\hspace{0.2cm} Equation~(\ref{eq:24}) with charge densities~(\ref{eq:25}) and~(\ref{eq:27}) can easily be solved and after some manipulations the results can be written as:
\begin{eqnarray*}
V_{I}(z)= -\frac{2\pi e^{2}}{\epsilon} \left\{ \begin{array}{ll}
-N_{S}z-2(N_{acc}+N_{S})w    & \big(z < -w\big) \\
(2N_{acc}+N_{S})z            & \big(-w < z < 0\big) \\
N_{S}z+2N_{acc}z(1-z/2l_{A}) & \big(0 < z < l_{A}\big)\\
N_{S}z+N_{acc}l_{A}          & \big(z > l_{A}\big)
\end{array} \right.
\end{eqnarray*}
\begin{eqnarray}
\lefteqn{ V_{S}(z) = \frac{2\pi e^{2}}{\epsilon}N_{S}[z-{}}
\nonumber\\
& & {}-2z\int^{\infty}_{0}f^{2}(z_{1})dz_{1}-2\int^{z}_{0}(z_{1}-z)f^{2}(z_{1})dz_{1}].
\nonumber\\
\label{eq:28}
\end{eqnarray}
where $N_{acc} = N_{D}l_{A}$ is the areal density of charged impurities in the accumulation layer. In formula~(\ref{eq:28}) we used condition~(\ref{eq:26}) and neglected for $z < -w$ the component of potential $V_{I}(z)$ square in z (which has a small effect on subband energies since the hole envelope functions vanish rapidly in the barrier).
The potential value at the end of accumulation layer:
\begin{eqnarray}
V(l_{A})\approx-\frac{2\pi e^{2}}{\epsilon}N_{acc}l_{A}-\frac{4\pi e^{2}}{\epsilon}N_{S}\int^{\infty}_{0}z_{1}f^{2}(z_{1})dz_{1}
\nonumber\\
\label{eq:29}
\end{eqnarray}
defines the position of valence band edge far from the interface (in the flat-band region) and is related to the hole Fermi energy $E_{F}$. If we denote by $E_{A}$ the position of acceptor levels outside the accumulation layer and by $\epsilon_{A}$ the bulk acceptor binding energy, we can write (see Fig.~\ref{fig:1}) $V(l_{A}) = E_{A} + \epsilon_{A}$. Solving the neutrality condition
\begin{eqnarray*}
N_{D}=N_{A}\Big[1+2\exp(\frac{E_{A}-E_{F}}{k_{B}T})\Big]^{-1}
\end{eqnarray*}
with respect to $E_{A}$ we obtain
\begin{eqnarray}
V(l_{A})=E_{F}+\epsilon_{A}+k_{B}T\ln \Big(\frac{N_{A}-N_{D}}{2N_{D}}\Big).
\label{eq:30}
\end{eqnarray}
Equations~(\ref{eq:29}) and~(\ref{eq:30}) allow us to calculate both the length $l_{A}$ and density $N_{acc}$ of the accumulation layer. The Fermi energy $E_{F}$ is determined from the known concentration of 2D holes. \\
\hspace*{0.2cm} Since the potential $V_{S}(z)$ produced by 2D carriers depends on the hole envelope function, the subband calculations must be self-consistent. At the first step we compute the potential $V(z)$ and the energy and wave function of the highest heavy-hole subband edge. After any iteration the Fermi energy $E_{F}$ is determined and new values of $l_{A}$ and $N_{acc}$ are calculated. If the in-plane wave vector $\vec{k}_{\bot}$ is equal to zero, all off-diagonal elements of Hamiltonian~(\ref{eq:9}) vanish. Therefore the subband edges can be calculated exactly by solving the Schr{\"o}dinger equation with scalar Hamiltonian~(\ref{eq:1}) and effective masses $m_{0}/(\gamma_{1}-2\gamma_{2})$ for heavy and $m_{0}/(\gamma_{1}+2\gamma_{2})$ for light holes. The Luttinger parameters used in the calculation are given in Table~\ref{tab:table1}. The value of valence-band offset $V_{0}=219.2$ meV is obtained as $35\%$ of the difference in band gaps in two adjacent layers. The concentration of residual donors $N_{D}$ is taken to be $5{\times}10^{14}$ cm$^{-3}$.
\begin{table}
\caption{\label{tab:table1} The valence-band parametres used in the calculation (after Ref. \cite{Landolt38}). Linear interpolation is applied for Al$_{x}$Ga$_{1-x}$As.}
\begin{ruledtabular}
\begin{tabular}{ccc}
               & GaAs & AlAs \\ \hline
  $\gamma_{1}$ & 6.85 & 3.45 \\
  $\gamma_{2}$ & 2.10 & 0.68 \\
 	$\gamma_{3}$ & 2.90 & 1.29 \\
 	$\kappa$     & 1.2  & 0.12 \\ 
\end{tabular}
\end{ruledtabular}
\end{table}
The concentration of residual donors $N_{D}$ is taken to be $5\times10^{14}$ cm$^{-3}$.
 \\
\hspace*{0.2cm} Figure~\ref{fig:1} shows the computed valence band profile and positions of the first three subband edges for the sample with hole density $N_{S} = 7.6{\times}10^{11}$ cm$^{-2}$. Our results can be directly compared with those obtained in Ref. \cite{Kolok5}, where hole levels in the same structure were calculated assuming the \textit{n}-type residual GaAs doping. In our case the electric field in the well region is substantially (almost two times) weaker. As a consequence we obtain higher values of hole subband edges and smaller intersubband distances (HH1 = $-38.1$ meV and LH1 = $-45.9$ meV as compared with values HH1 $\approx$ $-42$ meV and LH1 $\approx$ $-56$ meV reported in \cite{Kolok5}). In spite of a large 2D hole density only the ground HH1 subband is occupied and the Fermi level is located about $3$ meV on top of the next LH1 subband.
\begin{figure}
\includegraphics[width=8.6cm, height=10cm]{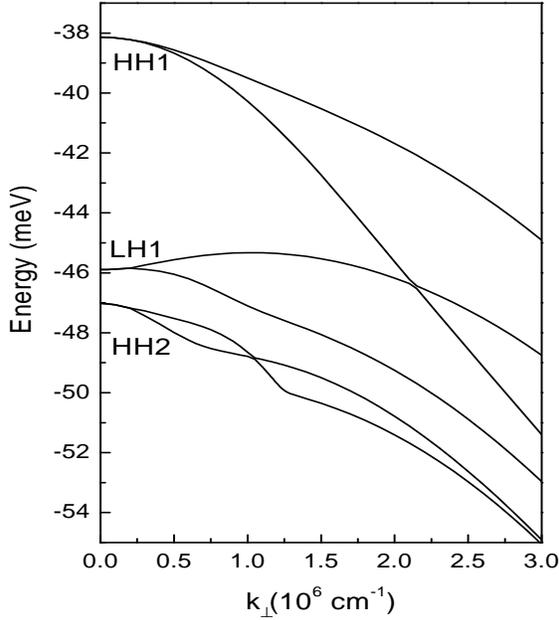}
\caption{\label{fig:2} Energy bands as a function of the wave vector $\vec{k}_{\bot}$ perpendicular to the growth direction. The sample parameters are the same as in Fig.~\ref{fig:1}.}
\end{figure} \\
\hspace*{0.2cm} Using the determined self-consistent potential $V(z)$ we next solve the Schr{\"o}dinger equation $(H_{L}+V)F = EF$ and calculate the subband dispersions for $\vec{k}_{\bot}\neq 0$. Results are plotted on Fig.~\ref{fig:2}. The subbands are spin-split (as a consequence of asymmetry of the confining potential) and strongly nonparabolic.  A comparison with the results of Ref. \cite{Kolok5} reveals that the difference in potential profile affects also the in-plane hole masses. For example a hole in the upper spin-split subband HH1 is in our model almost $30\%$ lighter.\\
\hspace*{0.2cm} Calculations including the magnetic field are performed using the same self-consistent potential as for $B = 0$. We therefore neglect any influence of the magnetic field on the band bending. The results will be discussed in next section. \\
\hspace*{0.2cm} To examine measured photoluminescence spectra we need to derive selection rules for interband optical transitions. The complete hole wave function for a given Landau level $n$ can be written as
\begin{eqnarray}
\Psi^{n}_{h}=F_{1}\phi_{n-1}u_{1}+F_{2}\phi_{n}u_{2}+F_{3}\phi_{n+1}u_{3}+F_{4}\phi_{n+2}u_{4}.
\nonumber\\
\label{eq:31}
\end{eqnarray}
where $u_{l}$ $(l=1,{\ldots},4)$ is the periodic part of the \textit{l}th Bloch function. The basis for
Hamiltonian~(\ref{eq:9}) is \cite{Andr35}
\begin{eqnarray}
u_{1} = \frac{1}{\sqrt{2}}(X+iY)\uparrow,
u_{2} = \frac{i}{\sqrt{6}}\Big[(X+iY)\downarrow - 2Z\uparrow\Big]\nonumber\\      
u_{3} = \frac{i}{\sqrt{6}}\Big[(X-iY)\uparrow + 2Z\downarrow\Big],
u_{4} = \frac{1}{\sqrt{2}}(X-iY)\downarrow.                                               \nonumber\\ 
\label{eq:32}
\end{eqnarray}
Similarly the total wave function of a conduction band electron has the form 
$\Psi^{e}_{n}=F_{e}(z)\phi_{n}u_{e}$ where $u_{e}=S\uparrow$ or $u_{e}=S\downarrow$  and the envelope function $F_{e}(z)$ represents an unbound motion. We calculate the matrix element $\left\langle\Psi^{h}_{n} \left|\vec{\varepsilon}\vec{p}\right|\Psi^{e}_{n'}\right\rangle$
for the transition in which the light with polarization $\vec{\varepsilon}$ is absorbed and an electron is raised from a state 
$\Psi^{h}_{n}$ to $\Psi^{e}_{n'}$. Since among matrix elements 
$\left\langle S\left|p_{v}\right|X\right\rangle$, $\left\langle S\left|p_{v}\right|Y\right\rangle$, $\left\langle S\left|p_{v}\right|Z\right\rangle$ 
($v=x,y,z$) merely
\begin{eqnarray}
\left\langle S\left|p_{v}\right|X\right\rangle=\left\langle S\left|p_{v}\right|Y\right\rangle=\left\langle S\left|p_{v}\right|Z\right\rangle=im_{o}P,
\label{eq:33}
\end{eqnarray}
are different from zero ($P$ being the Kane momentum matrix element), we obtain
\begin{eqnarray}
\left\langle\Psi^{h}_{n} \left|\vec{\varepsilon}\vec{p}\right|F_{e}(z)\phi_{n'}S\uparrow \right\rangle =     
-im_{o}P \Big[ \varepsilon^{-}\int{F^{\ast}_{1}\phi^{\ast}_{n-1}F_{e}\phi_{n'}d^{3}\vec{r}}
\nonumber \\
+\frac{1}{\sqrt{3}}\varepsilon^{+}\int{F^{\ast}_{3}\phi^{\ast}_{n+1}F_{e}\phi_{n'}d^{3}\vec{r}}
+i\sqrt{\frac{2}{3}}\varepsilon_{z} \int{F^{\ast}_{2}\phi^{\ast}_{n}F_{e}\phi_{n'}d^{3}\vec{r}} \Big]\nonumber\\
\label{eq:34}
\end{eqnarray}
and
\begin{eqnarray}
\left\langle\Psi^{h}_{n} \left|\vec{\varepsilon}\vec{p}\right|F_{e}(z)\phi_{n'}S\downarrow \right\rangle =     
-m_{o}P \Big[ \frac{1}{\sqrt{3}}\varepsilon^{-}\int{F^{\ast}_{2}\phi^{\ast}_{n}F_{e}\phi_{n'}d^{3}\vec{r}}
\nonumber \\
+\varepsilon^{+}\int{F^{\ast}_{4}\phi^{\ast}_{n+2}F_{e}\phi_{n'}d^{3}\vec{r}}
+i\sqrt{\frac{2}{3}}\varepsilon_{z} \int{F^{\ast}_{3}\phi^{\ast}_{n+1}F_{e}\phi_{n'}d^{3}\vec{r}} \Big],\nonumber\\
\label{eq:35}
\end{eqnarray}
where $\varepsilon^{\pm}=(\varepsilon_{x}\pm i\varepsilon_{y})/\sqrt{2}$. In our \textit{p}-doped structures the number of photoexcited electrons is small and only the lowest conduction-band Landau levels with $n'=0$ participate to the luminescence. An inspection of formulas~(\ref{eq:34})  and~(\ref{eq:35}) gives the selection rules that are summarized in Table~\ref{tab:table2}. We consider only optical transitions in the Faraday configuration.
\begin{table}
\caption{\label{tab:table2} Matrix elements of allowed interband transitions involving the lowest conduction-band Lnadau level for two different circular polarizations.}
\begin{tabular}{|c|c|c|c|} \hline \hline
 \multicolumn{1}{|c|}{polarization} & 
 \multicolumn{1}{c|}{hole} &
 \multicolumn{1}{c|}{electron} &
 \multicolumn{1}{c|}{matrix} \\
 														&
 \multicolumn{1}{c|}{Landau} &
 \multicolumn{1}{c|}{spin} &
 \multicolumn{1}{c|}{element} \\
 														&
 \multicolumn{1}{c|}{level} &
 														& \\ \hline
$\sigma^{-}$ & $n=1$  & $\uparrow$   & $P\int{F^{\ast}_{1}(z)F_{e}(z)dz}$ \\ \cline{2-4}
              & $n=0$  & $\downarrow$ & $(P/\sqrt{3})\int{F^{\ast}_{2}(z)F_{e}(z)dz}$ \\ \hline
 $\sigma^{+}$ & $n=-1$ & $\uparrow$   & $(P/\sqrt{3})\int{F^{\ast}_{3}(z)F_{e}(z)dz}$ \\ \cline{2-4}
              & $n=-2$  & $\downarrow$ & $P\int{F^{\ast}_{4}(z)F_{e}(z)dz}$ \\ \hline \hline
\end{tabular}
\end{table}
The transitions involving valence band Landau levels $n=0$ and $n=-1$ are three times less intense than the others, i.e. they have light-hole character. One can expect that the strongest transition takes place between the hole level $n=-2$ and the spin-down electron state in the $\sigma^{+}$ polarization. All the components of hole wave function~(\ref{eq:21}) different than $F_{4}(z)$ (the active component according to Table~\ref{tab:table2}) are then equal to zero.

\section{RESULTS AND DISCUSSION}
\begin{figure}
\includegraphics[width=9.7cm, height=7cm]{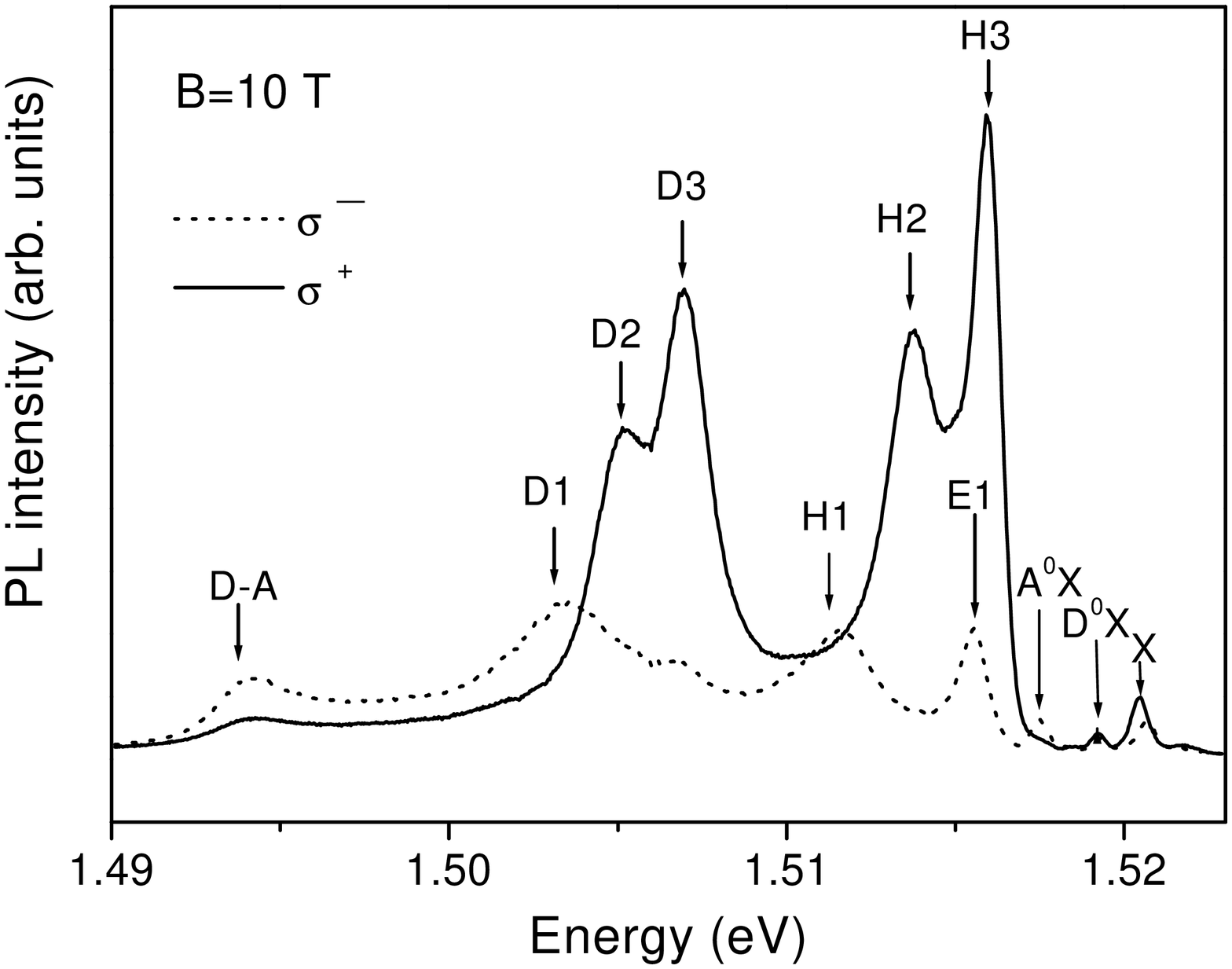}
\caption{\label{fig:3} Photoluminescence from sample with $N_{S} = 7.6{\times}10^{11}$ cm$^{-2}$ measured at $B=10$ T in different  circular polarizations for the excitation power density of $0.16$ mW/cm$^{2}$. The lines indicated by arrows are identified in the text.}
\end{figure} 

\begin{figure}
\includegraphics[width=8.6cm, height=10cm]{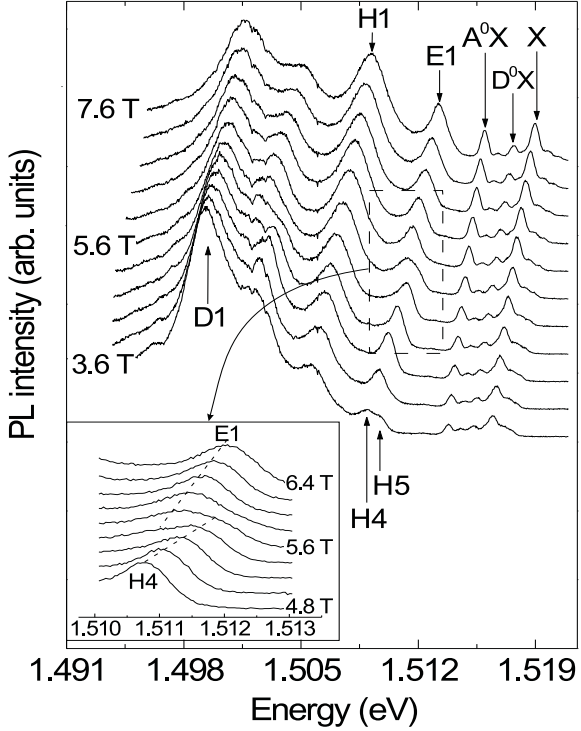}
\caption{\label{fig:6} Detailed PL spectra obtained in low magnetic fields in the $\sigma^{-}$ polarization. Near $B = 5.6$ T the H4 line involving $n=0$ holes declines and is replaced by a stronger peak with lower energy denoted as E1. At the same fields a high-energy donor transition (presumably related to $n=0$ holes) also disappears.}
\end{figure}

\begin{figure}
\includegraphics[width=9.5cm, height=7cm]{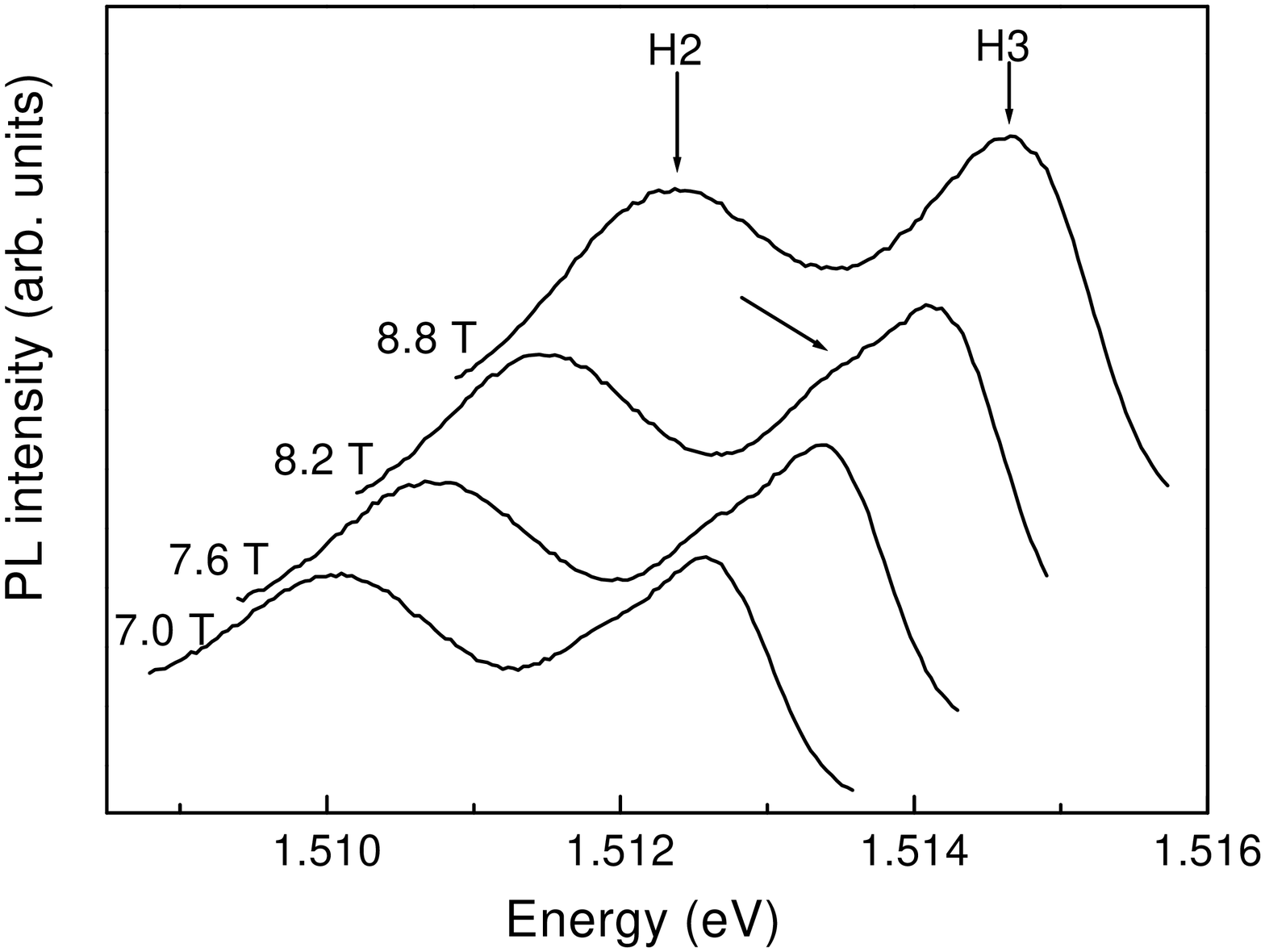}
\caption{\label{fig:7} Parts of PL spectra in the $\sigma^{+}$ polarization including free-hole lines measured near the crossing of $n=-1$ and $n=3$ valence-band levels. Due to wave-function mixing holes from the level $n=3$ can recombine with $n=0$ electrons and a weak peak (indicated by an arrow) appears on the low-energy shoulder of H3 transition.}
\end{figure}

\begin{figure}
\includegraphics[width=8.6cm, height=12cm]{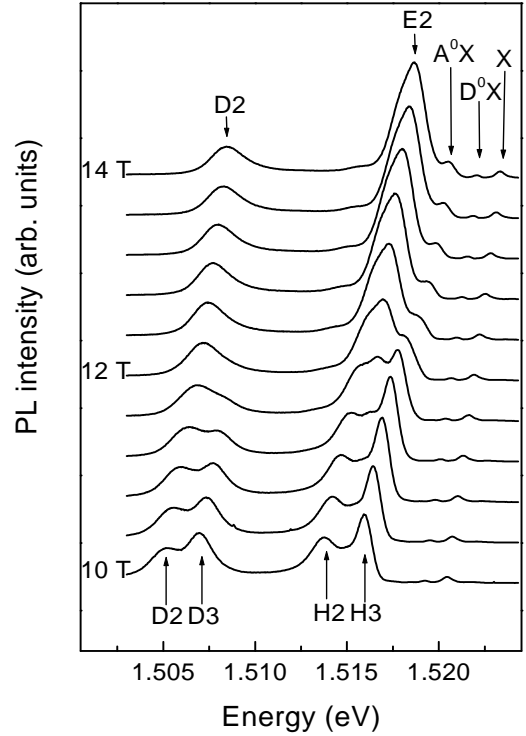}
\caption{\label{fig:8} High-field PL spectra obtained in the $\sigma^{+}$ polarization. The lines H3 and D3 related to the  $n=-1$ hole Landau level disappear at $B=11.6$ T and at the same field a new peak named E2 becomes visible. In still higher fields the new transition gains in intensity and masks the line H2 related to $n=-2$ holes. However the donor transition D2 also involving $n=-2$ holes remains visible.}
\end{figure}

\begin{figure}
\includegraphics[width=9.5cm, height=7cm]{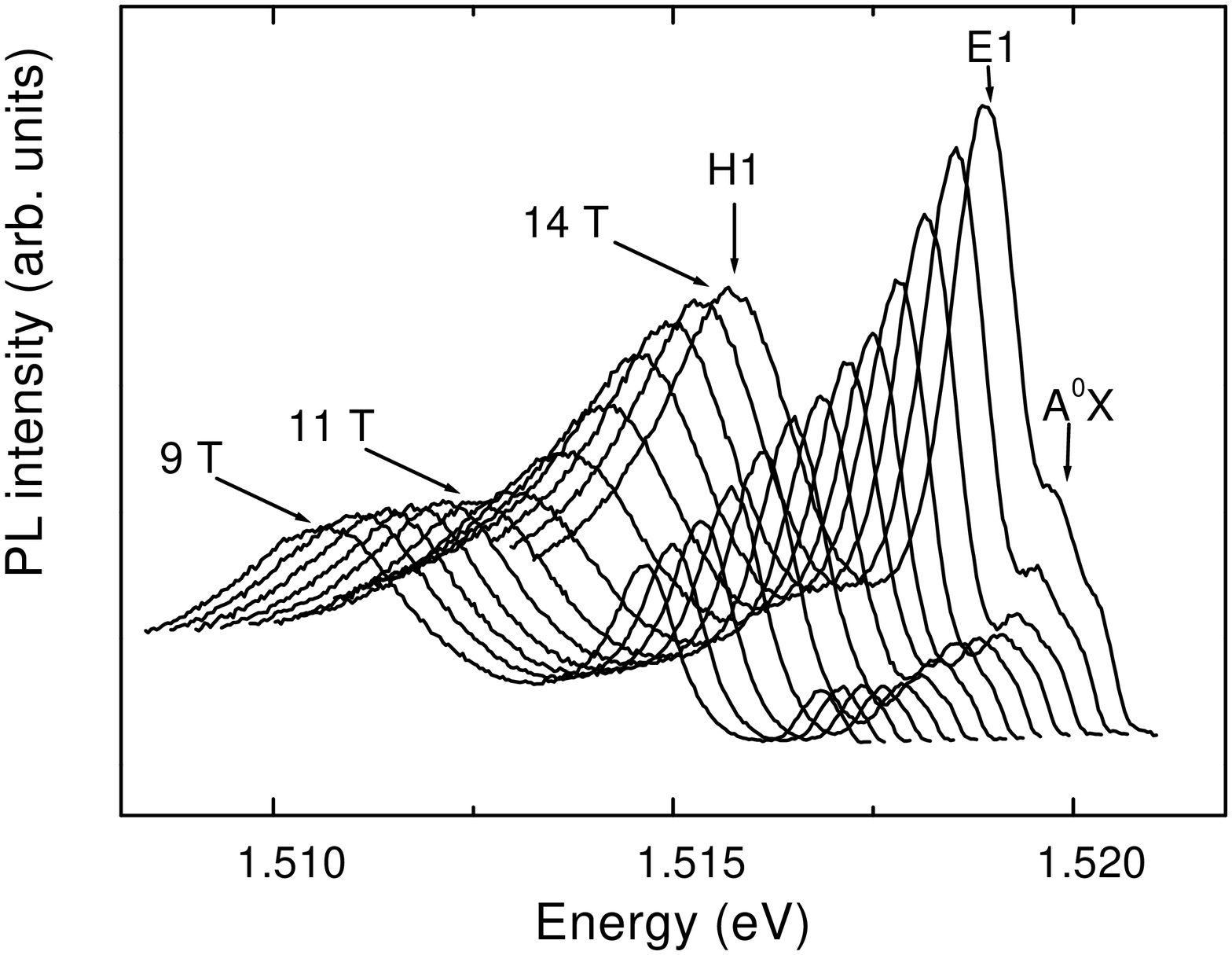}
\caption{\label{fig:9} Fragments of high-field PL spectra measured in the $\sigma^{-}$ polarization. As the free-carrier line H1 shifts to higher energies its intensity increases.}
\end{figure}

Photoluminescence spectra from the sample with $N_{S} = 7.6\times10^{11}$ cm$^{-2}$ measured at $B = 10$ T in both $\sigma^{+}$ and $\sigma^{-}$  polarizations are shown in Fig.~\ref{fig:3}.  
\begin{figure}
\includegraphics[width=8.6cm, height=12cm]{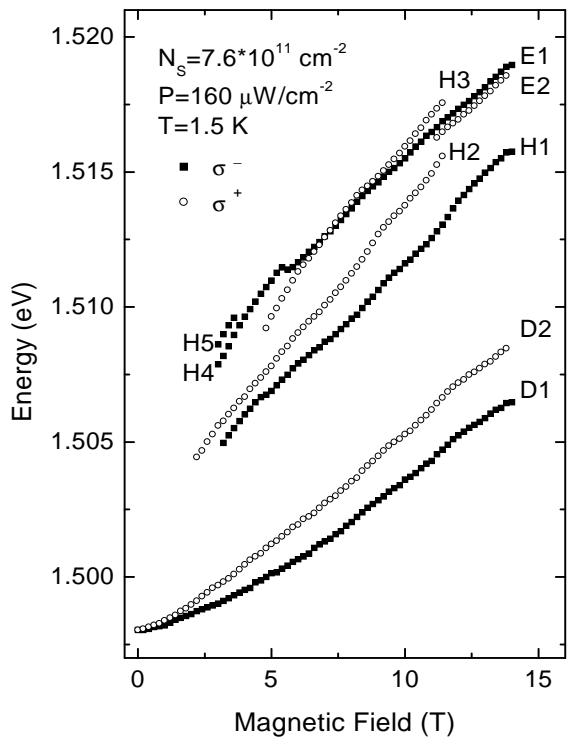}
\caption{\label{fig:4} Shift and splitting of selected photoluminescence lines in a magnetic field applied perpendicularly to the interface.}
\end{figure}
At low excitation powers most of the incident light is absorbed in the GaAs region near to the interface. This explains relatively small intensities of the bulk lines named X, D$^{0}$X and A$^{0}$X and attributed to free, donor-bound and acceptor-bound GaAs excitons, respectively. Stronger peaks denoted as H and D have two-dimensional character. This appears from the lines behavior with changed excitation power examined in our previous paper \cite{Bryja31}. We found that the relative intensities of X-lines increased as the laser power was raised and more of the incident light reached the flat-band GaAs region far from the interface. At the same time the X-line positions remained unchanged whereas those of both H and D peaks shifted towards higher energies. This results from the generation of electron-hole pairs which screen the electric field inside the accumulation layer. In effect the band bending is reduced and the energy of electron near the junction increases. On the other hand recombination processes in the flat-band region remain unaffected. The line named D-A arises from the donor-acceptor recombinations inside the accumulation layer. Both its position and intensity also depend on the laser power. \\
\hspace*{0.2cm}	We identified the H lines as resulting from the recombination of photoexcited electrons with 2D holes confined close to the junction. The lines named D were recognized as originating from the recombination of 2D holes with electrons bound to residual donors \cite{Bryja31}. Positions of all free-carrier transitions and of two donor lines with lowest energies as function of the magnetic field are plotted in Fig.~\ref{fig:4} (the lines denoted as E will be discussed later). They display different behavior: the D-lines exhibit diamagnetic shift whereas the energies of free-electron transitions change almost linearly with \textit{B} \cite{Inform}. At low fields the donor peaks are red-shifted with respect to the H-lines by the value of $4$ meV close to the bulk GaAs Rydberg energy and the difference increases with growing magnetic field. \\
\hspace*{0.2cm} To attain a high resolution the PL measurements were performed using very low density of excitation power $P = 0.16$ mW/cm$^{2}$. We observed that at excitation levels below $0.7$ mW/cm$^{2}$ the positions of H-lines saturate and do not change with the power \textit{P}. Presumably the number of photocreated carriers is so small that the band profiles remain unaltered. Under these conditions the donor-related emission is distinct even at zero magnetic field but the H-lines emerge only in higher fields $B>2$ T. With increasing excitation power the donor lines gradually disappear \cite{Bryja31}. This can be related to the known effect of saturation of a defect-related photoluminescence \cite{Bastard39}. \\
\hspace*{0.2cm} In this paper we analyze only the free-carrier transitions in an attempt to reveal the structure of valence-band states. To compare experimental results with calculated hole Landau levels we subtracted the measured transition energies from the energy $E_{cv} + 0.5\hbar\omega_{c}\pm0.5\mu_{B}gB$ of a GaAs conduction band electron in the lowest Landau level. Here $\mu_{B}$ is the Bohr magneton, g = $-0.44$ is the electron g-factor and the effective mass $m_{e}=0.067m_{0}$ was used to calculate the cyclotron frequency $\omega_{c}$. For each transition the projection of electron's spin was resolved according to Table II. The energy $E_{cv}$ was used as a fitting parameter and a good overall agreement was achieved for the value $E_{cv}=1464$ meV (which will be discussed later). 
\begin{figure*}
\includegraphics[width=14cm, height=9.5cm]{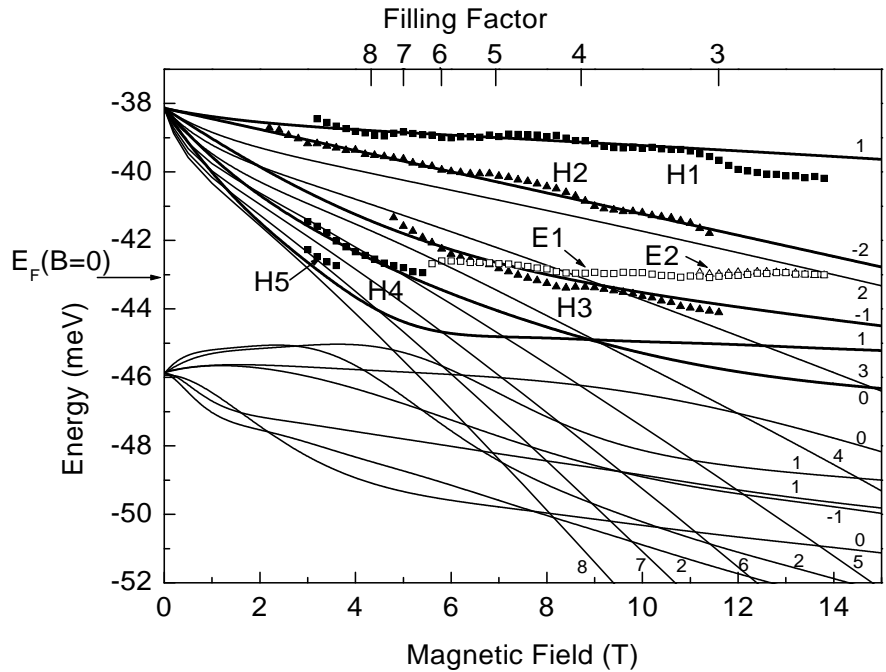}
\caption{\label{fig:5} Hole Landau levels in the sample with N$_{S} = 7.6{\times}10^{11}$ cm$^{-2}$ calculated (lines) and measured experimentally (symbols). In the ground subband only the levels are shown which influence observed transitions and optically active hole states are distinguished by thick lines. Squares and triangles represent emission lines detected in the $\sigma^{-}$ and $\sigma^{+}$ light polarization, respectively. Free-carrier transitions are indicated by full symbols. The filling factor was evaluated assuming somewhat higher 2D hole density of $8.4{\times}10^{11}$ cm$^{-2}$ as is discussed in the text.}
\end{figure*}
\\
\hspace*{0.2cm} The results are shown in Fig.~\ref{fig:5}. We plotted merely the positions of Landau levels with highest energies in the valence subbands HH1 and LH1. The $n=-2$ level appears only in the subband HH1 while Landau levels with $n\geqslant-1$ can be found in both subbands and for $n\geqslant+1$ there are two levels with the same number n in each subband ($n=0$ level emerges once in the subband HH1 and twice in LH1). At lower magnetic fields all the transitions displayed in Table II are observed. They successively disappear with increasing field as the involved hole Landau levels cross the Fermi level. The line H5 related to a hole from the lower $n=1$ Landau level (in the HH1 subband) vanishes at $B\approx 3.5$ T. This magnetic field value corresponds to the filling factor $\nu = 10$ at which the hole level becomes empty.\\
\hspace*{0.2cm} The next line H4 observed in the $\sigma^{-}$ polarization involves a $n=0$ hole. It disappears somewhat below $B=6$ T and simultaneously a new line, denoted as E1, emerges in the at slightly lower energy. This can be seen in detail in Fig.~\ref{fig:6} where we show the PL spectra recorded at fields near $6$ T. The $n=0$ level should be occupied by holes for filling factors $\nu\leqslant6$, i.e. up to $B=5.2$ T as calculated using the nominal 2D holes density. However the H4 line vanishes only at $B\approx 5.8$ T (see Fig.~\ref{fig:6}). This can be explained assuming that the actual hole concentration $N_{S}\approx 8.4\times 10^{11}$ cm $^{-2}$ exceeds the nominal value as an effect of sample illumination. 
 \\
\hspace*{0.2cm} According to Fig.~\ref{fig:5} the line H3 observed in the $\sigma^{+}$ polarization corresponds to the $n = -1$ hole Landau level. In the field range $2$ T $\leqslant B \leqslant 9$ T this is the fifth level (counting from the HH1 subband edge) and thus the H3 emission should disappear for filling factor $\nu = 4$ (i.e. $B = 8.7$ T). Indeed the peak intensity starts to fall at $B \approx 8$ T but for $B > 9$ T grows again and the line persists up to much higher fields $B > 11$ T. Furthermore as we can see in Fig.~\ref{fig:7} about $B \approx 8.2$ T the broadening of H3 line temporarily increases and even an additional peak emerges on the low-energy shoulder. This can easily be explained as a result of the crossing of $n=-1$ and $n=3$ levels predicted by the calculations (Fig.~\ref{fig:5}). It was shown \cite{Bang25} that cubic terms in Hamiltonian~(\ref{eq:9}) neglected in the axial approximation lead to strong Landau-level repulsion and to anticrossing behavior if the numbers \textit{n} of levels differ by $4$. This probably accounts for a slight discrepancy between calculated and measured energies of the $n = -1$ level near the crossing noticeable in Fig.~\ref{fig:5}. \\
\hspace*{0.2cm} In magnetic fields where the H3 transition disappears a new line labeled as E2 emerges in the $\sigma^{+}$ emission and rapidly gains in intensity. Its energy is greater than that of line E1 by the value $\mu_{B}gB$ of electron spin splitting. The detailed PL spectra are shown in Fig.~\ref{fig:8}. The energy of H3 transition can not be determined accurately for $B > 11$ T because it merges with the acceptor-bound exciton line $A^{0}X$. However we can find the field value at which the $n = -1$ hole level becomes depopulated by examining the donor-related lines D2 and D3. They correspond to recombinations of donor-bound electrons with $n = -2$ and  $n = -1$ holes, respectively. The line D3 disappears at $B \approx 11.6$ T, which agrees well with the value $\nu = 3$ calculated using the hole density $N_{S} = 8.4\times10^{11}$ cm$^{-2}$. At higher magnetic fields the peak H2 can not be separated from the strong E2 line. For this reason the H2 energies are not plot in Fig.~\ref{fig:5}. However the donor line D2 related to the same hole level $n = -2$, remains visible up to the highest magnetic fields. 
\\
\hspace*{0.2cm} Measured energies of PL lines (in particular of low-energy transitions H1 and H2) oscillate with the magnetic field. Similar behavior was frequently observed in n-doped structures \cite{Kim30,Skol40,Skol40A,Skol40B} and explained theoretically as a result of many-body interactions \cite{Kata41}. As in Ref. \cite{Kim30} we find that the amplitude of oscillations grows with B and the maximal and minimal values of transition energies correspond to even and odd filling factors, respectively. However in our \textit{p}-doped structures the intensity of luminescence increases at even filling factors, in contrast to what is reported in Ref. \cite{Kim30}. This can be seen in Fig.~\ref{fig:9} where we plot the lineshape of H1 transition in the fields range $9$ T $< B < 14$ T. The maximum of transition energy ($\nu = 2$) corresponds to $B = 15.7$ T above the limit of fields attainable in our experiment. To our knowledge the origin of magnetic-field induced oscillations of photoluminescence intensity remains unresolved.
\\
\hspace*{0.2cm} Apart from the oscillations the agreement between calculated and measured energies of hole Landau levels is very good. Also the intensities of observed transitions are consistent with matrix elements shown in Table II. The strongest transition H2 has intensity on average $2.5-3$ times greater than the line H1. Furthermore evaluated intensities of light-hole type transitions (H3, H4) are approximately three times less than those of heavy-hole type ones (H1 and H2).\\
\hspace*{0.2cm} The origin of lines E1 and E2 is not clear. They represent in different circular polarizations the same optical transition not related to a free-hole recombination. It is striking that the transition energy as plotted in Fig.~\ref{fig:5} coincides with the zero-field position $E_{F}$ of the hole Fermi level. This suggests a Fermi-edge singularity. However due to small masses of minority carriers in our structure (electrons in the GaAs layer) a possible singularity should be quite broad (smeared over the range $(m_{h}/m_{e})E_{F}$ \cite{Ruck42}). We see sharp peaks instead (Figs.~\ref{fig:6}, \ref{fig:8} and ~\ref{fig:9}) of a width below 1 meV. Moreover, as was observed in \textit{n}-doped structures \cite{Rubio43}, in strong magnetic fields the Fermi-edge singularity appears not as a new emission line but as some increase of the Landau-level transition when the level energy approaches the Fermi level. The other possible explanation - a recombination of electrons with acceptor-bound holes - must also be excluded. The line related to this transition can be noticed in the spectra (see Fig.~\ref{fig:6}) as a weak feature at energy about $15$ meV below the E-lines.
\\
\hspace*{0.2cm} In our view narrow widths and high intensities of  the E-lines suggest their excitonic nature. Similar line was seen by Kim et al \cite{Kim30} in the PL emission from a n-doped single GaAs/GaAlAs heterojunction. The peak red-shifted with respect to the free-electron transition which emerged at $\nu = 2$ was also interpreted as an excitonic recombination. In our structures photoexcited electrons are removed by the interface electric field far from the junction. This effect as well as screening of Coulomb interaction by 2D holes prevents a formation of excitonic states. This explains an absence of exciton lines in the PL emission at low magnetic fields. Nevertheless they can emerge at higher fields as the strong magnetic field applied in the growth direction brings an electron closer to a hole in the interface plane and increases excitonic binding. The resonance condition between states at the Fermi level and those of the excitonic level possibly further increases the intensity of E-line emission \cite{Muell44,Muell44A,Muell44B}. It is still not clear why the excitonic peak appears in much higher magnetic fields in the $\sigma^{+}$ than in the $\sigma^{-}$ polarization. As we have already noticed the E-line emerges only when a free-hole transition disappears. One can try to answer this question using the concept of phase-space filling. According to the theory \cite{Klei45} a hole Landau level can contribute to the excitonic wave function as soon as it becomes depopulated by carriers. Both levels $n=-2$ and $n=-1$ in the HH1 subband optically active in the $\sigma^{+}$ polarization are occupied by holes down to the filling factor $\nu = 3$. At higher magnetic fields the empty $n = -1$ states add to the excitonic level which can then produce also the $\sigma^{+}$ radiation. We intend further investigations, both experimental and theoretical, to clarify this effect.
\begin{figure*}
\includegraphics[width=14cm, height=9.5cm]{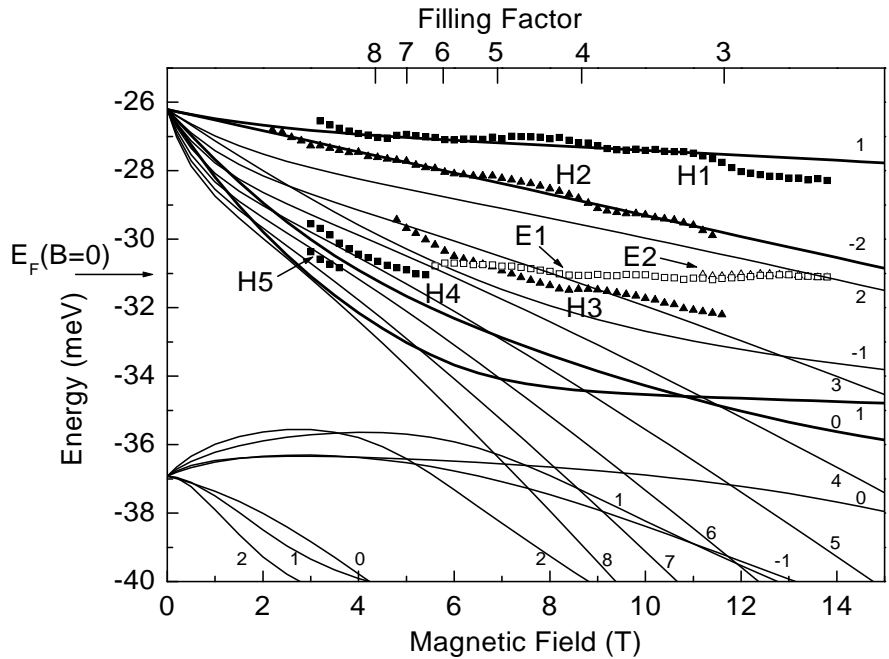}
\caption{\label{fig:10} Same as Fig.~\ref{fig:5} , with exchange-correlation corrections included in the calculations.}
\end{figure*}
\\
\hspace*{0.2cm} In strongly doped structures with high densities of free 2D carriers many-body effects beyond the Hartree approximation affect subband positions. It was shown theoretically \cite{Stern37} that in \textit{n}-doped heterojunctions corrections to electron energies due to exchange-correlation effects are quite small. However they should be more significant in \textit{p}-type systems because of large hole effective masses (in the local-density approximation the exchange-correlation potential is proportional to the effective Rydberg \cite{Stern37}). In previous calculations of hole Landau levels the many-body corrections were neglected. Their theoretical description is difficult due to complex structure of the valence band edge. A solution of this problem by means of the local-density approximation was proposed by Bobbert et al. \cite{Bobb26} and then successfully compared with the results of zero-field PL measurements on Be $\delta$-doped GaAs/Ga$_{1-x}$Al$_{x}$As quantum wells \cite{Kemer46}. We examined the influence of many-body effects on Landau level energies in our structures by adding the exchange-correlation potential from Ref. \cite{Bobb26} to the Hartree potential $V(z)$ during self-consistent calculations. The results are shown in Fig.~\ref{fig:10}. A comparison with Fig.~\ref{fig:5} reveals that exchange-correlation effects change significantly the subband edges. The ground HH1 subband is shifted by $12$ meV to higher energies and as a result the effective band gap is reduced. This effect is known as a band gap renormalization. Furthermore the splitting between the HH1 and LH1 subbands is increased by about $3$ meV. To compare calculated Landau levels with experimental results we once more subtracted the energies of measured transitions from the energy $E_{cv} + 0.5\hbar\omega_{c}{\pm}0.5\mu_{B}gB$, this time using the value $E_{cv}$ $12$ meV greater. As is seen the agreement remains very good for the highest levels $n=1$ and $n=-2$ but is substantially worse for the others. We verified that the fit can not be improved by small changes of the hole density value  used in calculations. However the new value of energy $E_{cv}$ is more consistent with experimental results. An inspection of the conduction band profile in the GaAs layer (computed by adding the band gap energy $E_{g}=1519.5$ meV to the calculated potential $V(z)$ shown in Fig.~\ref{fig:10}) reveals that it has energy $1449.5$ meV in the flat-band region far from the interface. Hence the value $E_{cv}=1464$ meV corresponds to an electron placed at least $75$ nm away from the junction, too far to recombine with confined holes. Increasing the energy $E_{cv}$ by $12$ meV one shifts the electron to a reasonable distance of about $40$ nm \cite{Bryja31}. We thus conclude that the model of Bobbert et al. \cite{Bobb26} adequately describes exchange-correlation corrections to the subband edge but distorts the effective masses of holes in excited (low-energy) Landau levels.

\section{SUMMARY}
	We investigated photoluminescence from \textit{p}-doped single heterostructures at helium temperatures and in a magnetic field directed normally to the interfaces. Very low excitation powers applied in the measurements allowed to obtain rich spectra containing both free and bound-carrier transitions involving two-dimensional holes. The free-carrier lines were used to derive the structure of valence-band Landau levels in confined subbands. We calculated energies and wave functions of hole states using a new numerical procedure based on a realistic model of charge distribution in studied heterojunctions. The theoretical results were consistent with experimental data. All allowed optical transitions between holes in occupied valence-band Landau levels and photocreated electrons from the ground conduction-band Landau level were detected. Their relative intensities were coherent with evaluated selection rules. The measured recombination energies agreed well with computed positions of hole levels. Also the field ranges in which individual transitions remained visible matched the calculated sequence of levels. Observation of Landau-level crossing provided information about the hole states which - due to symmetry constrains - cannot normally participate to the emission. Detailed comparison of experimental data with the theory allowed discerning the optical transitions (named as \textit{E}) with energies in the range of free-carrier recombinations but not matching the Landau-level scheme. Interestingly the E-line positions correlated with the hole Fermi energy in zero magnetic field. Moreover the line emerged in the $\sigma^{+}$ polarization in much higher magnetic fields than in the $\sigma^{-}$ one. We presented some arguments for the excitonic nature of related optical transition possibly influenced by a Fermi-edge singularity. Many-body interactions affected the PL spectra in different ways. We observed well-known oscillations of transition energies in a magnetic field correlated with integer values of the hole filling factor. Inclusion of exchange-correlation effects in Landau-level calculations along with the model recently proposed by Bobbert et al. \cite{Bobb26} gave mixed results. Computed correction to the ground-subband edge improved the agreement with experimental results but we obtained incorrect values of hole masses, particularly in excited Landau levels.

\begin{acknowledgments}
This work was supported by a grant from the Polish Committee for Scientific Research No.2 P03B08419.
\end{acknowledgments}

\newpage 

\begin{thebibliography}{53}
\bibitem[{Inf()}]{Inform}
\eprint{The different magnetic-field dependence of the D and H-lines is more
  evident when higher excitation powers are used in measurements [33]. This
  allows to observe free-carrier recombinations (H-lines) at lower magnetic
  fields.}
\end{thebibliography}

\end{document}